# SDR Common View on Generic Signals for Time Transfer Applications

Fabrizio Pollastri
INRIM
f.pollastri@inrim.it

*Abstract*—This paper presents a new time transfer system that works with any radio signal with sufficient bandwidth, regardless of its content and modulation, by adopting the common view approach. This system, based on a network client-server architecture with SDR receivers, offers a number of advantages. It can compare remote atomic clocks or disseminate reference time scales to end users with precision at the level of tens of nanoseconds. Its improved features in terms of flexibility, robustness, reliability, and security will potentially make positive contributions in the field of time transfer, as an alternative or complement to existing methods.

## I. INTRODUCTION

Timing is becoming increasingly important in various fields of science, production, and society, where technological systems need to be synchronized and reference time scales need to be disseminated to end users to improve overall performance, robustness, and security, along with the added value of traceability to the international time scale UTC. Over the years, several time transfer techniques based on different communication means, such as RF signals (mainly from GNSS satellites) and LAN/Internet networks, have been developed to fulfill these requirements. More recently, other technologies, like those based on optical fibers, have been developed to better meet specific technological requirements. At the same time, further research/development is being carried out to find alternative or complementary solutions in the field of time transfer.

One of the oldest techniques for time transfer is based on broadcast TV radio signals. The methods of this type changed over time to follow the technological evolution of TV radio signals from analog transmission [1] to the latest digital standards [2]. Both types of methods have one common feature: they exploit the modulation structure of the TV signal. For this reason, each method only works with the TV standard for which it was developed.

The main motivation for the time transfer system presented in this paper is to overcome the dependency on TV signal standards and, also, to be transparent to the radio signals used and resistant to jamming or spoofing attacks. With such a system, any type of radio signal with sufficient bandwidth can be used, regardless of its content and modulation. In addition, this system is based on a client-server architecture that communicates via the internet. The server is connected to a reference clock or a time scale, while one or more clients allow remote clocks to be compared/synchronized with the server reference, as well as to disseminate the reference time to end users.

The system is based on the common view approach, in which both the server and the clients receive a radio signal sent by the same transmitter. The server extracts a small piece of the signal and provides it with its reference timestamp. This information is sent to the clients as a time marker. In the meantime, the clients store the received signal and provide it with a timestamp that refers to the client's internal clock. When they receive a time marker, they search for the marker in the stored signal. If it is found, each client can compare its timestamp, which corresponds to the position of the stored signal where the time marker was found, with the timestamp of the time marker generated by the server and measure the time difference between the reference clock and the client's clock.

The time transfer system described above is well suited to exploit the capabilities of S*oftware Defined Radio* (SDR). The basic idea is to use a professional SDR for the server and low-cost open source SDR for the clients. In this work, a special effort was addressed to enhance an open source low-cost SDR, to become a customized and dedicated time transfer client unit, that can be used for a wide a range of timing applications.

In general, the systems based on SDRs [4] have a time resolution limit determined by the SDR sampling rate. The proposed system overcomes this limit by applying an efficient time resolution enhancement technique that allows for arbitrarily small instrumental time resolution. This technique is described in detail in Section VI . Thereby the main limit of the time resolution is determined by the signal quality which is affected by noise and multi-path effects.

## II. SYSTEM ARCHITECTURE

The functional principle of GSCV (Generic Signal Common View), the time transfer system presented here, can be better understood by referring to the diagram in Fig. 1.

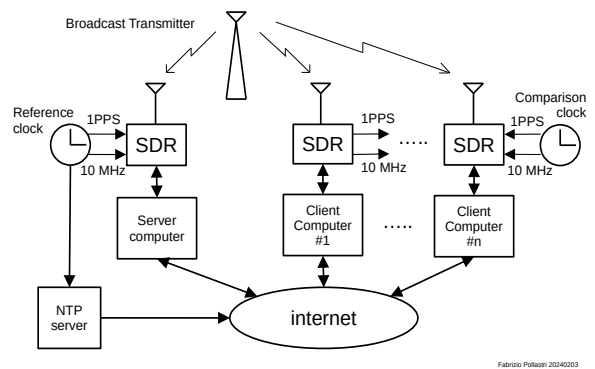

*Figure 1: system architecture diagram.*

A reference clock synchronizes a NTP server and a professional SDR via its 1PPS and 10 MHz inputs. The system server and all clients are roughly synchronized by the NTP server via the Internet. This coarse synchronization enables the clients to limit the search for the time marker to a time window of the received signal that is proportional to the uncertainty of the NTP synchronization. Typical values of 20-40 ms can be assumed for this uncertainty. Since the client must receive a time marker from the server for each search cycle, the search cycles are currently limited to 1 cycle/s in order to reduce network traffic and the required computational effort. The search cycles end with an estimate of the time offset between the clocks of the SDR server and the clients, so that a remote end user can compare/synchronize its clock with the reference one. Alternatively, such an estimated time offset can used to discipline the SDR client's internal clock, to remotely reproduce a replica of the server's reference clock, in the form of 1PPS and 10 MHz signals.

The requirements of a reference server SDR and clock comparator client SDRs are better met by professional SDRs, as these are normally equipped with inputs for synchronization signals. The synchronizer client SDRs can also be low-cost SDRs, which usually do not have 1PPS and 10Mhz synchronization signal inputs.

The implementation of GSCV presented in this paper is based on a client of the Synchronizer type. The client is implemented with a low-cost SDR, which is extended to become a synchronization generator that outputs 1PPs and 10MHz signals.

The network communication between the system server and the clients is based on UDP packets. Each client sends a "Send time marker" request to the server. The server then begins to send a time marker every second until a certain expiry time of the client is reached. Before this time, the client must send a new "Send time marker" request, if it wishes to maintain the transmission of time markers.

III. SDRs

This section describes the SDRs selected to set up a real working system with the architecture described above.

A. Server SDR

Following the basic idea of using a professional SDR for the server, an Ettus B200 was selected (Fig. 2b). This SDR can be synchronized in frequency with a 10 MHz input reference and in time with a 1PPS input. In addition, it offers full support for timed commands which are essential for accurate time stamping of the sampled signal. Full software support for program applications is freely available in both C++ and Python languages.

Its main features are:

- maximum sampling rate 61.44 MS/s
- 12 bit sampling resolution
- I/Q analytical sampling
- bandwidth range 200 kHz – 56 MHz
- tuning range 70 MHz – 6 GHz
- direct conversion to baseband

It has a USB 3 interface for host communication. This interface can lead to speed limitations with hosts that only have USB 2 interfaces. In practice, you can use a USB 2 for a sampling rate of up to 10 MS/s in this system.

B. Client SDR

On the client side, a HackRF One was chosen because it is low-cost and completely open in both software and hardware (Fig. 2c). In addition, this SDR board was designed to interface with other add-on boards, so it has numerous connectors that allow easy access to many of the internal signals. This is a key factor in expanding its capabilities, so that it is fully customizable as a dedicated client for time transfer applications. This SDR can be frequency synchronized with a 10 MHz input reference, but it lacks a 1PPS input for time synchronization and there is also no support for timed commands. These shortcomings were addressed by developing a new version of the SDR firmware and utilizing the built-in expansion flexibility of the SDR hardware. There is an open source software support consisting of a C language library and a set of tools for interacting with the SDR. There is also an open source Python wrapper of the C library, developed by others, but only partially implements the library's API.

Its main features are:

- maximum sampling rate 20 MS/s
- 8 bit sampling resolution
- I/Q analytical sampling
- bandwidth range 1 MHz – 15 MHz
- tuning range 1 MHz – 6 GHz
- direct conversion to baseband

It has a USB 2 interface for host communication. The practical application in this system, allows you to use a USB 2 for a sampling rate of up to 10 MS/s.

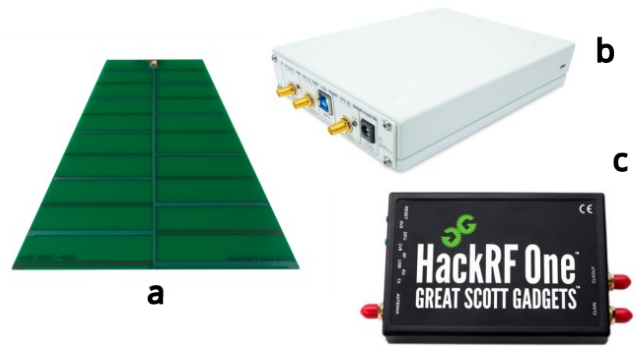

*Figure 2: a) log periodic UHF antenna, b) Ettus B200 SDR, c) HackRF One SDR.*

IV. FIRMWARE AND SOFTWARE DESIGN

The system architecture shown in Fig. 1 requires the development of two application programs: one for the server and one for the clients. An additional design criterion is that the language used should be at the highest level and compatible with both SDR software development packages. It should also run on a Linux operating system in order to achieve a completely open source solution. With these assumptions, the Python language is the best choice.

A first preliminary step, is to define the capabilities required for the application programs.

*A.* Server program: required capabilities

- Periodic extraction (at least one per second) of the time marker from the received signal as "common view".
- Validation of the time marker based on specific criteria.
- Management of a list of active clients waiting to receive time markers from the server.
- Transmission of the time marker to the clients.

*B.* Client program: required capabilities

- Periodic (at least once per second) signal acquisition in the time marker search windows.
- Request and reception of time markers.
- Search for the received time marker in the acquired search windows.
- If a search results in a valid match, measure the phase offset between client and server time markers.
- Use the measured offset to adjust the client SDR's clock.

*C.* Client SDR firmware: required capabilities

Another important part of the system that requires profound changes is the client SDR. Fortunately, due to its high hardware accessibility (see III.B) and the architecture of its functional blocks, this model is suitable to be customized and used as a low-cost GSCV client, potentially addressing mass-marked users. The HackRF One's microcontroller clock can be switched from its normal fixed frequency source, a 12 MHz crystal, to a programmable clock generator: the same clock generator provides the sampling clock. So if you set both the sampling frequency and the microcontroller clock to 10 MHz, you can set the generated frequency with a resolution of 0.25 Hz (25 ns). The rest of the resources needed for this project are in the microcontroller itself. It has a lot of free peripherals, in particular two unused timers/counters with counting, dividing and pulse generation capabilities, with the pulse outputs available on the expansion connectors.

Given these available hardware resources, the firmware modification requirements are as follows:

- Configure a microcontroller timer to produce a 1PPS pulse.
- Establish a firmware counter for tracking seconds.
- Configure the timer to generate a trigger pulse with programmable delay relative to the 1PPS pulse for precise sampling start time.
- Enable fine adjustments for sampling and timer clocks.

*D.* Server program implementation

Figure 3 shows a block diagram of the server program. The program runs in two separate processes to enhance real-time response. The first process, named "signal process", handles all I/O operations with the SDR and the extraction and validation of the time marker. An SDR driver module interfaces with the SDR, mapping the interface functions specific to the type of SDR, to a virtual SDR with a standardized interface. Currently, two drivers have been developed: one for the Ettus B200 and one for the HackRF One. The second process, named the "network process", manages all network-related operations: waiting for an incoming marker transmission request, maintaining the list of active clients, and transmitting markers to all active clients. All network operations are executed as concurrent tasks by asynchronous coroutines.

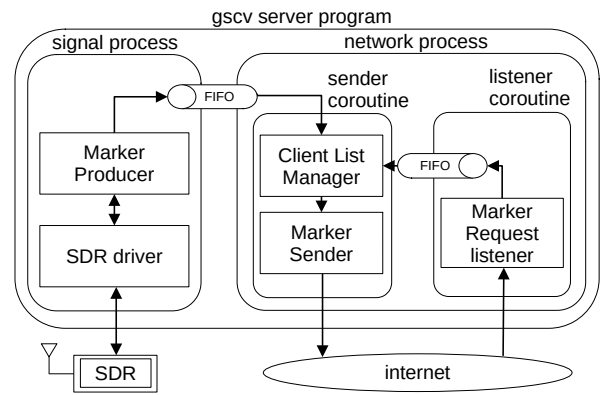

*Figure 3: server program diagram.*

*E.* Client program implementation

Figure 4 shows a block diagram of the client program. Similar to the server program, this program runs in two separate processes and interfaces with the SDR through a driver module. The process named "signal process" handles all the I/O operations with the SDR, including searching and validating the time marker, as well as adjusting the clock according to the time marker phase just measured. The other process, named "network process", is responsible for all network-related tasks such as receiving the time marker and periodically transmitting a marker request.

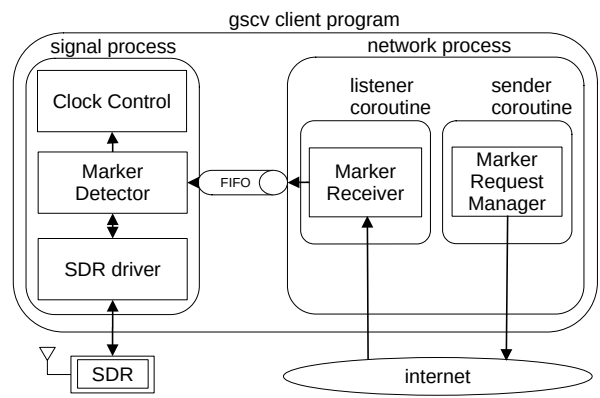

*Figure 4: client program diagram.*

*F. SDR firmware implementation*

Figure 5 shows the block diagram of the HackRF One extensions implemented in both the firmware and by utilizing the existing hardware resources. The following firmware functions have been added. Functions for complete control of the clock generator, as the existing ones do not allow full control. Functions to configure a spare timer of the MCU microcontroller as a time ticks counter. Functions to configure the generation of 1PPS pulse. Functions to configure the generation of a trigger pulse for precise sampling start.

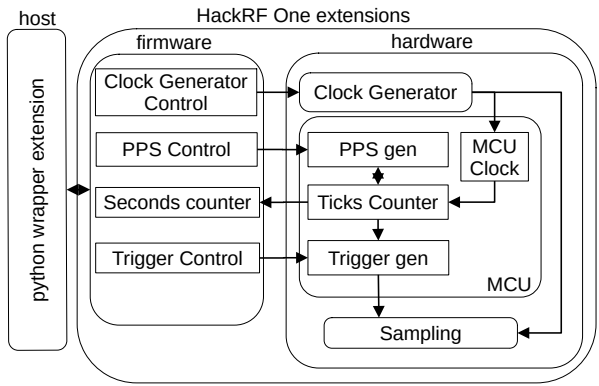

*Figure 5: HackRF One extensions diagram.*

The 1PPS pulse generated by the hardware triggers the increment of a second counter implemented by the extended firmware.

*G. Python wrapper*

Since the SDR firmware is written in the C language, a python wrapper is necessary to access the firmware functions from python application programs. A python wrapper for the Hackrf One SDR, called pyhackrf2, was previously developed [9]. However, it did not encompasses all the functions of the HackRF One. Therefore, for the GSCV system, it was enhanced with additional functions that were missing and expanded to encompass all the new functions introduced in the SDR firmware.

V. TIME MARKER

As illustrated above, the time marker, its detection and the assessment of its reliability are essential components of the GSCV system. Since this system purports to be fully transparent regarding the signal utilized for the common view time transfer, the primary method of establishing a time marker involves extracting a portion of the signal at the server SDR, transmitting it to the clients and then utilizing a correlation detector to locate it within the signal received by the client SDRs.

*A. Marker validation*

The output of the correlation detector typically shows multiple peaks. The peak with the highest value (indicating the maximum similarity between time marker and the received signal from the client) is considered the most suitable candidate for the marker position within the received signal. However, the requirement to ignore any signal characteristics can result in correlation ambiguities, where one or more peaks may have values very close to the candidate peak. To address this issue, a method is needed to validate the candidate peak against other peaks. The approach chosen involves calculating the relative height difference between the candidate peak and the highest of other peaks. If this difference exceeds a given threshold, the candidate peak is deemed "valid". Tests with the common view signal described in section IX, have shown that 0.3 is a practical value: it means that the second height peak is 30% lower than the candidate peak.

*B. Marker shaping*

Marker shaping or how to extract a portion of the common view signal received by the system server, results from a balance between two conflicting needs:

- marker uniqueness or the highest level of marker validity;
- minimal marker size for efficient network transmission.

Tests using the simplest marker shape, a single rectangular window of the common view signal, indicates sizes of approximately 4000 samples at a sample rate of 10 MHz, with the common view signal described in section IX, to achieve a high level of validity. However, this type of marker has a significant impact on network transmission.

Similarly, high levels of validity can be achieved with a much smaller marker in the form of a scattered sequence of small rectangular windows. For instance, 16 windows of 50 samples each, spread across 10000 samples interval. This is feasible because, when considering a signal with unknown characteristics, the likelihood of two samples correlating decreases as their temporal distance increases. This marker shape enables a reduction of the marker size by 5 times when compared with the single window shape.

The implementations of the correlation detector typically utilize the correlation theorem for efficient computation, involving both direct and inverse FFT (Fast Fourier Transform). The FFT assumes that the input signal is periodic, repeating outside the windows of definition. If the signal is non-zero at the window boundaries, false transitions occur, leading to artifacts in the frequency domain. Windows with fewer samples are more prone to this issue. To mitigate this, various window weighting functions have been proposed [11], all designed to smooth the signal to zero at the window edges. Therefore, for marker of this shape, the applying a suitable weighting window is essential before conducting the correlation detection.

*C. Marker windows position*

As the marker consist of a sequence of small sparse windows of signal samples, they are positioned at random distances from each other. This pattern offers two additional advantages.

- enhances the system resistance to spoofing;
- improves the validity rate of the marker by decreasing the correlation between the marker and periodic patterns in the signal.

*D. Marker search window*

As mentioned earlier, the client look for the time marker within a time window proportional to the NTP

synchronization uncertainty between the server and the client. When the client is on the same site as the server, the search window can be limited to a typical value of 10 ms. For a remote client, the search window is expanded to 40 ms. This search range is adequate for medium-quality network connections where the NTP can synchronize the client with an uncertainty of less than 10 ms.

## VI. TIME RESOLUTION

This system achieves a maximum intrinsic time resolution of 1 nanosecond (ns) in three stages.

Initially, during the client system startup, the phase difference of the time marker is measured relative to its nominal position in units of the sampling period (10 MHz). Subsequently, the next 1PPS cycle of the HackRF is adjusted by this difference to quickly and roughly reduce the marker phase difference.

During each client cycle, the phase difference is measured again taking the maximum of the correlation detector's modulus between the marker and the signal. This measure offers a resolution equivalent to the sampling period, which is 100 ns for a 10 MHz sampling rate.

Given that both the marker and the signal are band-limited, the correlation modulus is also band limited. Therefore, the Withaker-Shannon interpolation [6] can be utilized on the correlation modulus to achieve an exact signal reconstruction with the desired temporal resolution of 1 ns. This reconstruction is specifically applied to a narrow interval around the approximate maximum position found in the previous stage to enhance the maximum position from 100 ns to 1 ns resolution.

## VII. PID CONTROLLER

The phase offset, measured with 1 ns resolution, is feed to a PID controller [7]. Its output is used as a frequency difference (limited to +-10 Hz) which is then applied to the client sampling frequency and to the PPS counter frequency. Because the client error signal is a phase measure and the control signal is a frequency, the client can be represented as an integrator. Considering this, the dominant parameter of the PID is the proportional one. The integral and the derivative parameters where adjusted through several tests.

## VIII. SECURITY

The network communication between the server and the clients can be protected using a standard cryptographic protocol like DTLS [12]. The server identity is verified by a key pre-shared with clients. This key, along with the time of each cycle, is utilized to determine the random position of the marker windows. This approach adds an additional layer of anti-spoofing security.

Jamming threats can be more effectively addresses by the GSCV system compared to GNSS synchronization systems. The signals utilized by GSCV originate from terrestrial transmitters that are more powerful and closer to the client and server receivers than any satellite transmitter used in GNSS. Furthermore, the receiving antennas used in GNSS need to have wide receiving lobes, so they are sensitive to any jamming source in the sky. On the contrary, GSCV antennas can be highly directive and any jammer outside the trasmitter-receiver line of sight is greatly attenuated. Additionally, the number of broadcast transmitters suitable for GSCV is much higher than the number of GNSS satellites, providing a wide range of options for pointing the GSCV antennas.

## IX. SYSTEM TESTS

To preliminary evaluate the GSCV potentials and possible limitations, an experimental measurement setup was installed at the INRIM Radio Navigation Laboratory. With such a setup, the following functional tests have been performed:

a) for the evaluation of its system noise, the GSCV system was operated at the common clock and zero baseline configuration, connecting both the Server and the Client to the 10 MHz signals of UTC(IT) [10], the Italian Standard Time, as well as to the same antenna, using a dedicated RF splitter; this test revealed a precision limit of the GSCV system of approximately 1 ns standard deviation and 3 ns peak-to-peak noise;

b) to assess synchronization noise, the 10 MHz input reference to the client SDR is disconnected and the GSCV system client software was switched to synchronization mode. The client SDR clock frequency was regulated by a PID controller's output, which takes as input the difference between the position identified by searching the received marker in the common view signal and its extraction position at the server SDR. The remaining components of the previous setup a remain unchanged;

c) to evaluate the GSCV system under real operating conditions, the server setup from the previous setup b was kept unaltered, while the client was relocated 50 km away to a remote site and operated without an external clock reference in synchronization mode.

For all tests, a transmitter located in sight of view about 8 km away from the server, broadcasting a DVB-T2 multiplex at 626 MHz was chosen as the source of the common view signal. All tests were addressed to estimating the precision of the GSCV system.

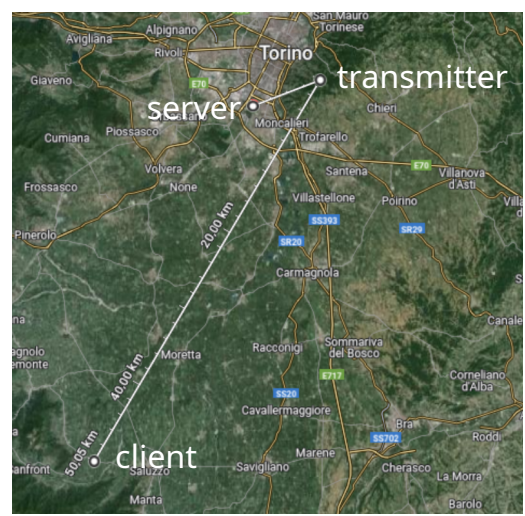

*Figure 6: map of client, server and transmitter for test with the client at a remote site.*

Figure 7 depicts a typical plot achievable with the zero baseline setup b, lasting 1000 seconds. The uppermost graph illustrates the phase offset between the client's time marker and the server-extracted marker (phase origin), disregarding any propagation delay. The plot displays a standard deviation of approximately 10 ns and a peak-to-peak noise of about 35 ns, showcasing a good performance in precision.

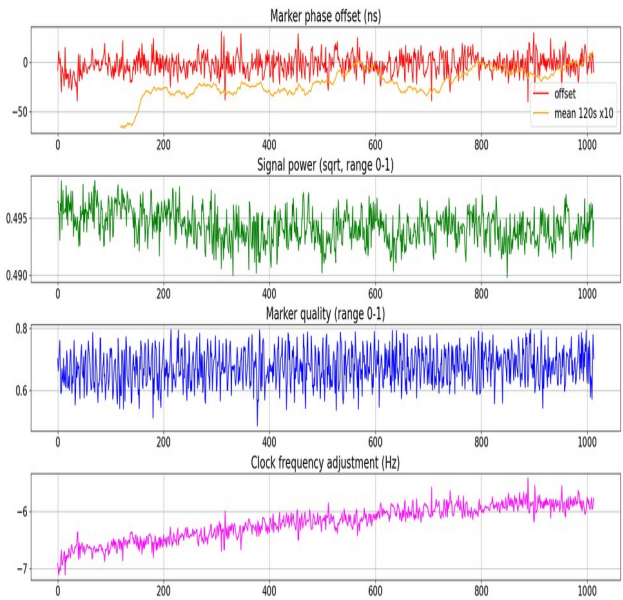

*Figure 7: typical marker phase, signal power, marker quality and frequency adjustment at the local client, setup b.*

The second graph depicts minor fluctuations of the signal power. The third graph illustrates the marker quality, which remains quite good with an average around 0.6 and a standard deviation of 0.05. The bottom graph displays the frequency adjustment implemented by client's control loop PID. The minimum acceptable threshold is established at 0.3.

Figure 8 shows a typical plot achievable with the remote client setup c. The plot shares similar characteristics with the zero baseline test, showing a standard deviation of approximately 10 ns and a peak-to-peak noise of 36 ns. The signal power fluctuations are more pronounced compared with the local client scenario, primarily due to signal multi-paths. However, the marker quality remains quite good with an average of about 0.7 and a standard deviation of 0.05.

Tests b and c demonstrate consistent precision in marker phase detection. This means that the increased distance of the client from the transmitter (from 8 to 50 km) does not significantly impact precision, with the main limitation being the client's SDR performance.

Throughout all conducted tests, the frequency adjustment plot displays a characteristic curvature resembling a warm-up thermal transient lasting approximately 1000 seconds when the HackRF board is exposed to free air, and around 2000 seconds when enclosed.

These findings affirm that GSCV system possesses instrumental limits suitable for synchronization and time transfer.

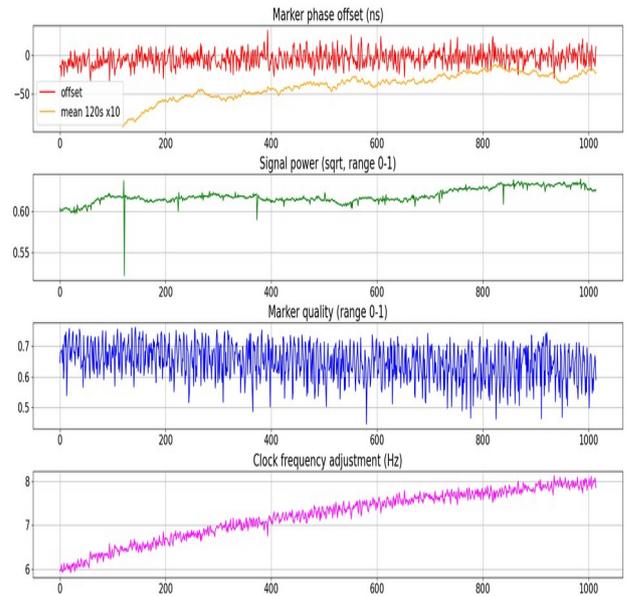

*Figure 8: typical marker phase, signal power, marker quality and frequency adjustment at the remote client, setup c.*

X. CONCLUSIONS

The GSCV time transfer system alternative to the traditional systems, such as those utilizing GNSS signals, is presented. It operates in common view mode with any radio signal that has sufficient bandwidth, providing a precision better than 40 ns and utilizing a low-cost modified SDR. These characteristics make it suitable for a wide range of applications requiring an alternative time transfer system.

Further research will focus on investigating the system's accuracy, testing new types of time markers and control loops. Additionally, different sources of radio signals from satellite television broadcasts will be considered to expand the system coverage area. To enhance jamming resistance, a frequency hopping mode for receiving signals will be explored.

To evaluate the system accuracy for absolute time transfer, measurements of all involved delays, propagation, cables and instruments will be conducted with metrological traceability.

ACKNOWLEDGMENTS AND DISCLAIMER

The GSCV system was developed as part of the project 'FRATERNISE: high temporal accurate facility for fundamental physics experiments' [13] founded by the Fondazione CRT (Cassa di Risparmio di Torino | RF= 2021.0893) and with liberal contributions from the Bank of Italy.

I want to express my gratitude to Elena Cantoni, Giancarlo Cerretto, and Marco Sellone at INRIM for their invaluable support.

The use of brand names and/or any mention or listing of specific commercial products is solely for scientific research purposes and does not imply endorsement by the author, nor discrimination against similar brands, products not mentioned.

## REFERENCES

[1] J. Tolman, V. Ptáček, A. Souček, and R. Stecer, "Microsecond Clock Comparison by Means of TV Synchronizing Pulses", IEEE Trans. on Instrumentation and Measurement, vol. IM-16, no. 3, sep. 1967.

[2] D. Boehm, J. White, S. Mitchell and E. Powers, "Clock Comparison Using Digital Television Signals", Proceedings of the 2009 Precise Time and Time Interval, pp. 319-326.

[3] K. Tarasov, B. Bauer, S. Olsen, E. Grayver, H. Feil, J. Sherman, A. Montare, M. Deutch, G. Nelson, M. Lombardi, T. Marczewski and D. Howe, "Time Transfer using High-Definition Television (HDTV) Broadcast Transmitters in Common View", Proceedings of the 2023 Precise Time and Time Interval, pp. 140-149.

[4] M. Bartolucci, J. A. Del Peral-Rosado, R. Estatuet-Castillo, J. A. Garcia-Molina, M. Crisci and G. E. Corazza, "Synchronisation of low-cost open source SDRs for navigation applications," 2016 8th ESA Workshop on Satellite Navigation Technologies and European Workshop on GNSS Signals and Signal Processing (NAVITEC), Noordwijk, Netherlands, 2016, pp. 1-7, doi: 10.1109/NAVITEC.2016.7849328.

[5] "HackRF One, an open source SDR platform", https://greatscottgadgets.com/hackrf/, [Online; accessed 25-Nov-2023].

[6] Claude E. Shannon, "Communication in the Presence of Noise", Proc. IEEE, vol. 86, no, 2, pp. 447-457, feb, 1998, https://web.archive.org/web/20100208112344/http://www.stanford.edu/class/ee104/shannonpaper.pdf, [Online; accessed 02-May-2024].

[7] Aidan O'Dwyer, "Handbook of PI and PID Controller Tuning Rules", 3rd ed. Imperial College Press. ISBN 978-1-84816-242-6.

[8] D. W. Allan, "Should the classical variance be used as a basic measure in standards metrology?," in IEEE Transactions on Instrumentation and Measurement, vol. IM-36, no. 2, pp. 646-654, June 1987, doi: 10.1109/TIM.1987.6312761.

[9] "A Python wrappper for libhackrf rewritten", https://pypi.org/project/pyhackrf2/, [Online; accessed 05-Feb-2024].

[10] "Improvements in the Realization of the Italian Time Scale UTC(IT)", https://hdl.handle.net/11696/75826, [Online; accessed 07-Feb-2024].

[11] "Understanding FFTs and Windowing", https://download.ni.com/evaluation/pxi/Understanding FFTs and Windowing.pdf, [Online; accessed 15-Feb-2024].

[12] "The Datagram Transport Layer Security (DTLS) Protocol Version 1.3", https://datatracker.ietf.org/doc/html/rfc9147, [Online; accessed 01-May-2024].

[13] G. Cerretto, E. Cantoni, M. Sellone, et al., "Time Metrology for Fundamental Physics Experiments: INRIM's Experience with the FRATERNISE Project", in Proceedings of the ION 2024 Precise Time and Time Interval Systems and Applications Meeting, January 22–25, 2024, Hyatt Regency Long Beach, Long Beach, California, ISBN: 978-0-936406-37-4.

■